\documentclass[a4paper,10.5pt,twocolumn,twoside]{article}

\usepackage{graphicx}
\usepackage{color}
\usepackage[a4paper, left=1.4cm, right=1cm, top=1.7cm, bottom=1cm]{geometry}
\usepackage[super,comma,sort&compress,square]{natbib}
\usepackage{amsmath}
\usepackage{float}
\usepackage{latexsym}
\usepackage{amssymb}

\usepackage{fancyhdr}
\usepackage[utf8x]{inputenc}
\usepackage[T1]{fontenc}

\begin{document}
\pagestyle{fancy}
\def\headrulewidth{0.5pt}
\def\footrulewidth{0pt}
\lhead{ACS Applied Materials \& Interfaces 12 (2020), 39926 -- 39934} 
\chead{}
\rhead{DOI: 10.1021/acsami.0c14115}

\lfoot{} 
\cfoot{}
\rfoot{}

\twocolumn[
  \begin{@twocolumnfalse}
  {\huge \bf Magnetization Reversal Mechanism in Exchange-Biased Spring-like Thin-Film Composite}

  \hspace{1.1cm}
  \parbox{.87\textwidth}{
    \vspace{4ex}
    \Large \textsf{Marcin Perzanowski, Jakub Gregor-Pawlowski, Arkadiusz Zarzycki, Marta~Marszalek}
    \vspace{1ex} \\
    \normalsize Institute of Nuclear Physics Polish Academy of Sciences, Deparment of Materials Science, Radzikowskiego 152, 31-342 Krakow, Poland  \\
    \vspace{-0.7ex} \\
    \normalsize \text{email: Marcin.Perzanowski@ifj.edu.pl}

    \vspace{1.5ex} 
    \noindent
     \textbf{Abstract}: Development of modern spintronic devices requires materials exhibiting specific magnetic effects. 
In this paper, we investigate a~magnetization reversal mechanism in a [Co/Pd$_{x}$]$_{7}$/CoO/[Co/Pd$_{y}$]$_{7}$ thin-film composite where an antiferromagnet is sandwiched between a~hard and a~soft ferromagnets with different coercivities. 
The antiferromagnet/ferromagnet interfaces give rise to the exchange bias effect. 
The application of soft and hard ferromagnetic films causes exchange-spring-like behavior while the choice of the Co/Pd multilayers provides large out-of-plane magnetic anisotropy. 
We observed that the magnitude and the sign of the exchange bias anisotropy field are related to the arrangement of the magnetic moments in the antiferromagnetic layer.  
This ordering is induced by the spin orientation present in neighboring ferromagnetic films which is, in turn, dependent on the orientation and strength of the external magnetic field. 
     
     \vspace{2ex}
     DOI: 10.1021/acsami.0c14115 (OPEN ACCESS)
     
     \vspace{2ex}
     
    \vspace{3ex}
  }
  \end{@twocolumnfalse}
]

\section{Introduction}

\vspace{-0.3cm}
Exchange bias effect is a phenomenon occurring at the interface between a~ferromagnet (FM) and an antiferromagnet (AFM). 
The exchange coupling occurs after cooling such a FM/AFM system in the external magnetic field below the N\'eel temperature of the AFM, giving rise to the magnetic hysteresis loop shift along the field axis.\cite{Kiw01JMMM,Nog99JMMM} 
The shift is called the exchange bias field $H_{\mathrm{ex}}$ and its magnitude is inversely proportional to the thickness of the FM material revealing the interfacial nature of the effect.  
In most cases the bias field decreases monotonically with increasing temperature to the field $H_{\mathrm{ex}}=0$ for the blocking temperature for exchange bias. However, there are also systems where the bias field first increases, and then drops down as temperature rises.\cite{Shi18JAP}
Usually, the blocking temperature is lower than the N\'eel temperature of the AFM due to the structural imperfections present in FM and AFM materials, as well as the condition of the interface. 
To describe the exchange bias phenomenon various models have been applied considering the magnetic domains in the antiferromagnet,\cite{Mil00PRL} the role of uncompensated spins \cite{Tak97PRL} as well as the roughness of the FM/AFM interface.\cite{Mal87PRB} 

The possible technological application of the exchange bias effect has been studied in the context of its implementation in sensors, \cite{Neg09APL} biomedicine, \cite{Ehr15S,Iss13IJMS} and magnetic read heads and spintronic devices.\cite{Par03IEEE,Gas13APL,Pol14APL} 
Most of the research on the exchange bias effect has been done on flat multilayers, however, there are also studies for materials in the form of magnetic antidots and dots,\cite{Per17AMI,Car14JAP,Suc09APL} core-shell structures,\cite{Sal16AMI,Swi18N,Shi14N} or rings and disks.\cite{Tri09N,Gil15N} 
One of the key features especially important for application in flat and patterned spintronic devices is a~perpendicular magnetic anisotropy present in the magnetic material. 
For this reason, we focused on system including Co/Pd ferromagnetic multilayers with easy axis of magnetization perpendicular to a~film plane.\cite{Car85APL,Car03APL} 
As an antiferromagnetic material for the exchange bias effect studies we have chosen cobalt oxide CoO since its properties are well known and the Co/CoO interface is considered to be a~model system for such investigations.\cite{Men14JAP,Per16AMI,Dob12PRB,Dia14JAP}  

Here, we present studies on the cooling field influence on the magnetization reversal mechanism for the exchange-biased system where the CoO antiferromagnetic layer is sandwiched between two [Co/Pd] ferromagnetic multilayers with different coercivities. 
The issue of the cooling field impact on the magnitude and sign of the exchange bias field has been recently raised in a~few research papers. However, these works focused on AFM/FM bilayer,\cite{Hu19N} spin glass/FM interface,\cite{Rui15SR} or system where AFM and FM layers are separated by a~paramagnetic material.\cite{Tor17N} 
This paper develops this field of research further by combining two magnetic effects in one study --- exchange bias and magnetic exchange spring, to find how they affect the reversal process. 
Such type of FM/AFM/FM composite that we study here is similar to the hard-soft exchange spring materials which can be applied in high density magnetic recording devices.\cite{Thi03APL,Ast06PRB,Sou10PRB} 
We find that the magnetization switching process and the magnitude and the sign of the exchange bias field are different depending on the magnetic state of both the ferromagnetic and the antiferromagnetic films induced by cooling the system in various external magnetic fields.
The studies of the magnetic hysteresis loops obtained under different conditions are supported by the First Order Reversal Curve (FORC) measurements. 
The FORC investigations were carried out for both ascending and descending branches of the hysteresis loop which is especially significant for the exchange-biased systems with bias loop shift from zero position. 

\vspace{-0.3cm}
\section{Experimental}

\vspace{-0.2cm}
The samples were fabricated by thermal evaporation at room temperature in ultrahigh vacuum under a~pressure of $10^{-7}$~Pa. 
The systems were deposited on single-crystal Si(100) substrates with a 2 nm thick Pd buffer layer. 
Ferromagnetic multilayers consisted of [Co/Pd]$_{7}$ stacks with a~Co thickness of 0.3~nm. 
The Pd layers have a~thickness of 0.6 or 1.2~nm. 
The antiferromagnetic CoO layer (AFM) was obtained by oxidizing 1-nm-thick Co layer in the atmosphere of pure oxygen under a pressure of $3 \times 10^{2}$~Pa for 10 minutes. 
In this paper four systems were studied --- the [Co/Pd$_{\mathrm{0.6 \ nm}}$]$_{7}$ and  [Co/Pd$_{\mathrm{1.2 \ nm}}$]$_{7}$ multilayers, the [Co/Pd$_{\mathrm{0.6 \ nm}}$]$_{7}$/[Co/Pd$_{\mathrm{1.2 \ nm}}$]$_{7}$ and the [Co/Pd$_{\mathrm{0.6 \ nm}}$]$_{7}$/CoO/[Co/Pd$_{\mathrm{1.2 \ nm}}$]$_{7}$ composites.
During the deposition the substrates were kept at 300 K.

The multilayer structure of the samples was investigated by X-ray reflectivity (XRR) measurements carried out using an X'Pert Pro PANalytical diffractometer equipped with a~Cu X-ray tube operated at 40~kV and 30~mA. 
For systems without AFM layer the XRR method was used only for validation of the assumed layered structure, while for AFM-based composite it was also applied to determine the CoO thickness and density.

Magnetic studies were done using Quantum Design MPMS XL SQUID magnetometer. 
The zero-field-cooled (ZFC) and field-cooled (FC) magnetization curves were measured as follows. 
First, the sample was demagnetized at 300 K by application of an osciltextitg external magnetic field, inducing zero magnetization. 
Then, the sample was cooled down to 10 K without the presence of the external field. 
At 10~K, the field of 500~Oe was applied in the direction perpendicular to the sample surface, and the ZFC magnetization curve was measured during heating to 300~K. 
After reaching the final temperature, the FC curve was recorded during cooling down to 10~K with the same external field. 
Both ZFC and FC curves were measured at a~temperature change rate of 3~K/min.
The hysteresis loops were measured in out-of-plane geometry at 10~K, 50~K, and 100~K, with the external magnetic field perpendicular to the sample plane. 
During the cooling down to low a~temperature various external magnetic cooling fields were applied. 
For more details see further in the text. 

First-order reversal curve (FORC) measurements were done at 10~K in out-of-plane geometry. 
During an FORC measurement along a~single hysteresis branch, first, a~$\pm$10~kOe field was set to magnetically saturate the ferromagnetic components of the system. 
Then, the field was changed to the reversal field $H_{\mathrm{R}}$ and the magnetization was measured as a~function of the external field $H$ toward the initial $\pm$10~kOe point. 
Next, the succeeding $H_{\mathrm{R}}$ field was set prior to the subsequent magnetization measurement.
The set of $M(H,H_{\mathrm{R}})$ curves measured for different starting $H_{\mathrm{R}}$ fields was transformed into an FORC distribution. 

\vspace{-0.3cm}
\section{Results and discussion}

\begin{figure*}[t]
 \centering
 \includegraphics[width=0.8\textwidth]{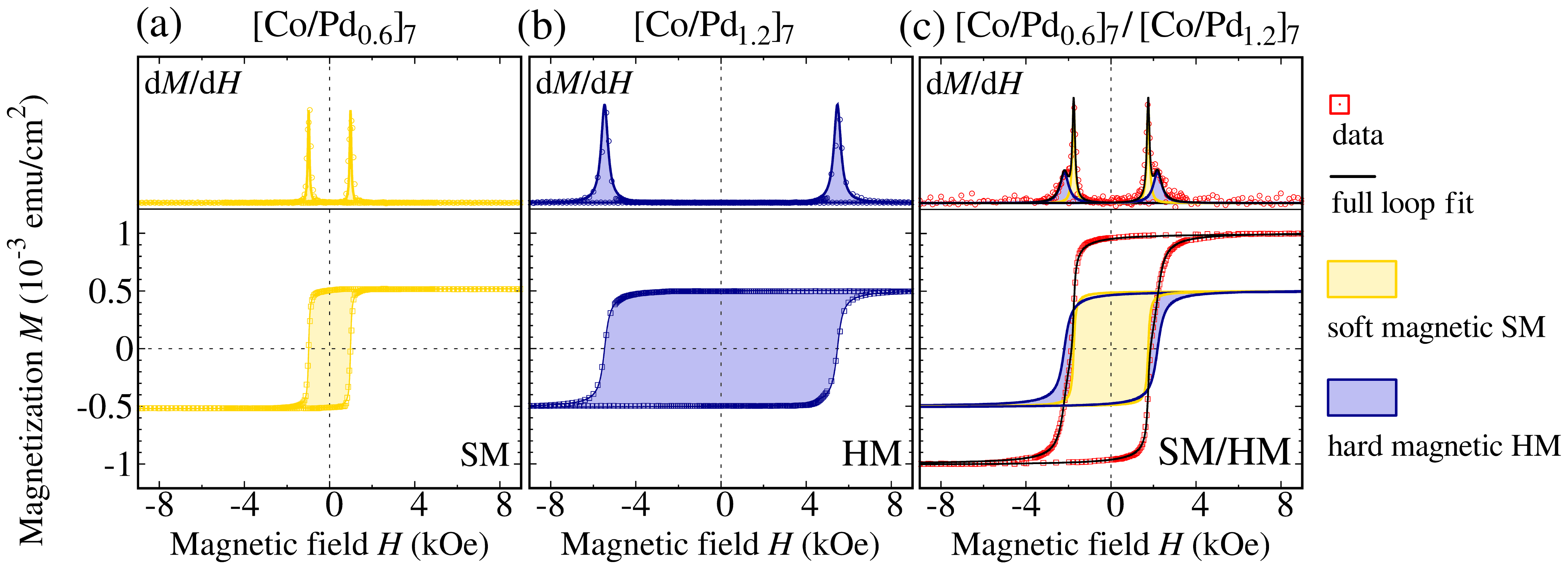}
 \caption{Magnetic hysteresis loops measured at 10~K in out-of-plane geometry for (a) [Co/Pd$_{\mathrm{0.6 \ nm}}$]$_{7}$, (b) [Co/Pd$_{\mathrm{1.2 \ nm}}$]$_{7}$, and (c) [Co/Pd$_{\mathrm{0.6 \ nm}}$]$_{7}$/[Co/Pd$_{\mathrm{1.2 \ nm}}$]$_{7}$ systems after cooling in a~+50~kOe external magnetic field. The upper panels present d$M$/d$H$ switching field distributions. The points represent experimental data, and solid lines are fits (see text). In Figure (c) the hard magnet (HM) and soft magnet (SM) components are marked with corresponding color fields.}
 \label{Fig1}
\end{figure*}

First, we studied the magnetic properties of two ferromagnetic components of a~complex composite system. 
The hysteresis loops together with the d$M$/d$H$ switching field distributions, measured at 10~K after field cooling in +50~kOe for [Co/Pd$_{\mathrm{0.6 \ nm}}$]$_{7}$ and [Co/Pd$_{\mathrm{1.2 \ nm}}$]$_{7}$ multilayers, and [Co/Pd$_{\mathrm{0.6 \ nm}}$]$_{7}$/[Co/Pd$_{\mathrm{1.2 \ nm}}$]$_{7}$ system, are presented in Figure~\ref{Fig1}. 
To obtain quantitative information on the magnetic properties of the systems both upper and lower magnetization branches of the hysteresis loops for the [Co/Pd$_{\mathrm{0.6 \ nm}}$]$_{7}$ and [Co/Pd$_{\mathrm{1.2 \ nm}}$]$_{7}$ multilayers were fitted using the following expression\cite{Per17AMI,Ste94JAP,Kuz99JMMM}
\begin{equation}
M(H) = \frac{2}{\pi} M_{\mathrm{s}} \arctan \left( g \left[ \frac{H-H_{\mathrm{c}}}{H_{\mathrm{c}}} \right] \right) \ ,
\label{single_M_H}
\end{equation}
where $M_{\mathrm{s}}$ is the saturation magnetization of the system, $H_{\mathrm{c}}$ is its coercitivy, and $g$ represents the slope of the magnetization curve.

The [Co/Pd$_{\mathrm{0.6 \ nm}}$]$_{7}$ multilayer (Figure~\ref{Fig1}a) shows remanence magnetization equal to 0.98 of saturation magnetization $M_{\mathrm{s}}$, indicating that the easy axis of magnetization is perpendicular to the sample plane. 
The switching field distributions d$M$/d$H$ for both upper and lower magnetization branches are symmetrical and sharp denoting an abrupt magnetization reversal process. 
The coercivity of the sample obtained from the fit is 1~kOe. 
The [Co/Pd$_{\mathrm{1.2 \ nm}}$]$_{7}$ multilayer (Figure~\ref{Fig1}b) has a~similar remanence demonstrating large out-of-plane magnetic anisotropy. 
In comparison to the [Co/Pd$_{\mathrm{0.6 \ nm}}$]$_{7}$ multilayer, the switching field distributions d$M$/d$H$ of this sample are broader showing that the rotation of the magnetic moments induced by the magnetic field sweep takes place in a more gradual way.
The coercivity of this multilayer is 5.4~kOe, and therefore, it will be denoted further in the text as a~hard magnet (HM), while the [Co/Pd$_{\mathrm{0.6 \ nm}}$]$_{7}$ multilayer will be labeled as a~soft magnet (SM). 
In both cases the same total Co thickness was deposited, and the saturation magnetizations normalized to the surface area are equal.
Due to the symmetrical single-step magnetization reversal both SM and HM systems can be described as rigid magnets. 

The hysteresis loop and the d$M$/d$H$ distributions for the [Co/Pd$_{\mathrm{0.6 \ nm}}$]$_{7}$/[Co/Pd$_{\mathrm{1.2 \ nm}}$]$_{7}$ system, being a~superposition of the hard and soft ferromagnets and further labeled as SM/HM, are shown in Figure~\ref{Fig1}c. 
The d$M$/d$H$ curves demonstrate asymmetrical switching field distributions related to the presence of different magnetic phases SM and HM reversing at distinct external magnetic fields.  
To acquire quantitive information on coercive fields and saturation magnetizations of the components each magnetization branch was fitted using a~sum of two $M(H)$ functions expressed by Eq.~\eqref{single_M_H}
\begin{equation}
\begin{split}
 M(H) = \frac{2}{\pi} M_{\mathrm{s}}^{\mathrm{SM}} \arctan \left( g^{\mathrm{SM}} \left[ \frac{H-H_{\mathrm{c}}^{\mathrm{SM}}}{H_{\mathrm{c}}^{\mathrm{SM}}} \right] \right) \\
 + \frac{2}{\pi} M_{\mathrm{s}}^{\mathrm{HM}} \arctan \left( g^{\mathrm{HM}} \left[ \frac{H-H_{\mathrm{c}}^{\mathrm{HM}}}{H_{\mathrm{c}}^{\mathrm{HM}}} \right] \right)  \ ,
\end{split}
 \label{double_M_H}
 \end{equation}
where the SM and HM superscripts denote the quantities associated with the corresponding constituent multilayers.

The sharper maxima, indicating swift magnetization reversal, occur for the magnetic field of $\pm1.7$~kOe and they are associated with the reversal of the SM component.
Accordingly, the broader d$M$/d$H$ maxima, suggesting slower process of magnetization rotation, giving a~coercivity of 2.2~kOe are related to the HM multilayer.
The saturation magnetizations of both SM and HM components are similar to those observed for the single [Co/Pd] multilayers (Figures~\ref{Fig1}a and~\ref{Fig1}b).
As a~consequence, the SM/HM system has the saturation magnetization twice as large as the individual [Co/Pd] stack.
The two-step reversal process demonstrates that the SM/HM composite acts like an exchange-spring system rather than like a~single rigid magnet.\cite{Kne91IEEE,Ful98PRB} 
The increase of the SM and the decrease of the HM coercivities, in comparison to the single [Co/Pd] stacks, result from the exchange coupling between magnetic materials. 
Similar changes observed for exchange-spring films with out-of-plane anisotropy were reported by Casoli et al.\cite{Cas08APL}

\begin{figure}[h]
 \centering
 \includegraphics[width=0.35\textwidth]{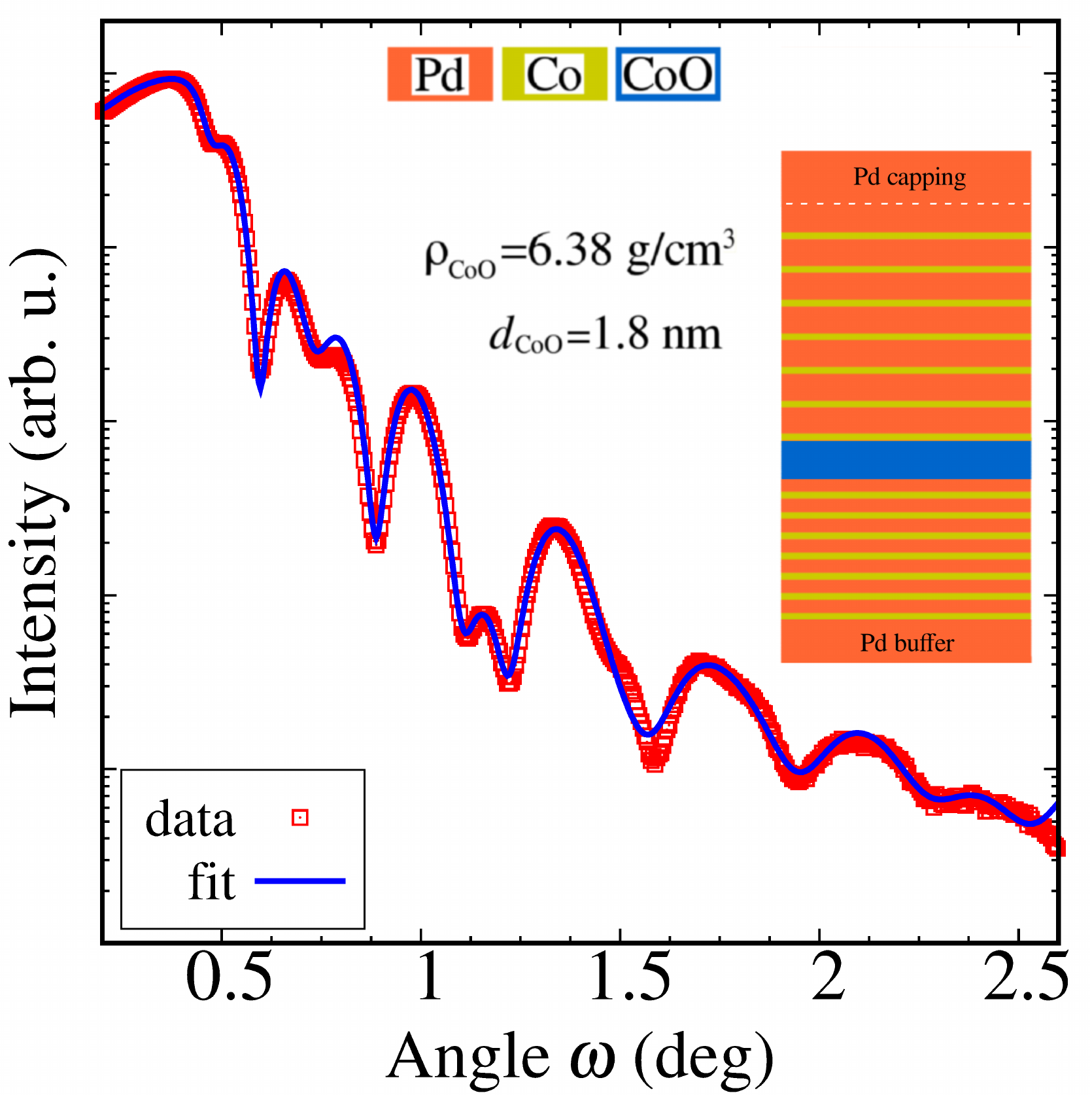}
 \caption{XRR measurement (red points) and fitted curve (blue line) for the [Co/Pd$_{\mathrm{0.6 \ nm}}$]$_{7}$/CoO/[Co/Pd$_{\mathrm{1.2 \ nm}}$]$_{7}$ composite.}
 \label{Fig2}
\end{figure}

\newpage
In order to study the magnetization reversal mechanism of a spring-like exchange-biased system, an antiferromagnetic (AFM) CoO layer was sandwiched between SM and HM multilayers. 
The XRR measurement (Figure~\ref{Fig2}) showed that the density of the cobalt oxide layer is 6.38~g/cm$^{3}$ which is close to the bulk CoO value of 6.44~g/cm$^{3}$. 
Therefore, we can expect that the Co layer was oxidized to CoO and it has appropriate magnetic properties to create magnetic exchange coupling with the ferromagnetic layers. 
The thickness of the CoO film determined from the XRR equals to 1.8~nm which is a~larger value than the deposited Co thickness due to the incorporation of atoms in the material. 
According to the work by van der Zaag et al.,\cite{Zaa00PRL} for such a~CoO thickness the blocking temperature for the exchange bias is approximately 160~K. 
Accordingly, the temperature of 10~K, at which magnetization reversal studies were carried out, is sufficiently low to register meaningful exchange bias field. 
The fit shows that the roughness of the Co layers within the [Co/Pd] stacks is comparable to their thickness equal to 0.3~nm, indicating large jaggedness of the Co-Pd interfaces. 
Therefore, it is highly unlikely that the deposited Co forms continuous layers. 
Taking into account the calculated roughness of each Pd layer being approximately 0.4~nm, the [Co/Pd$_{\mathrm{0.6 \ nm}}$]$_{7}$ stack should instead be considered as an intermixed Co-Pd film. 
In the case of the [Co/Pd$_{\mathrm{1.2 \ nm}}$]$_{7}$ multilayer, due to the larger thickness of the Pd layers, the intermixed Co-Pd interface regions are separated from each other by the layers of pure Pd.

\begin{figure}[t]
 \centering
 \includegraphics[width=0.35\textwidth]{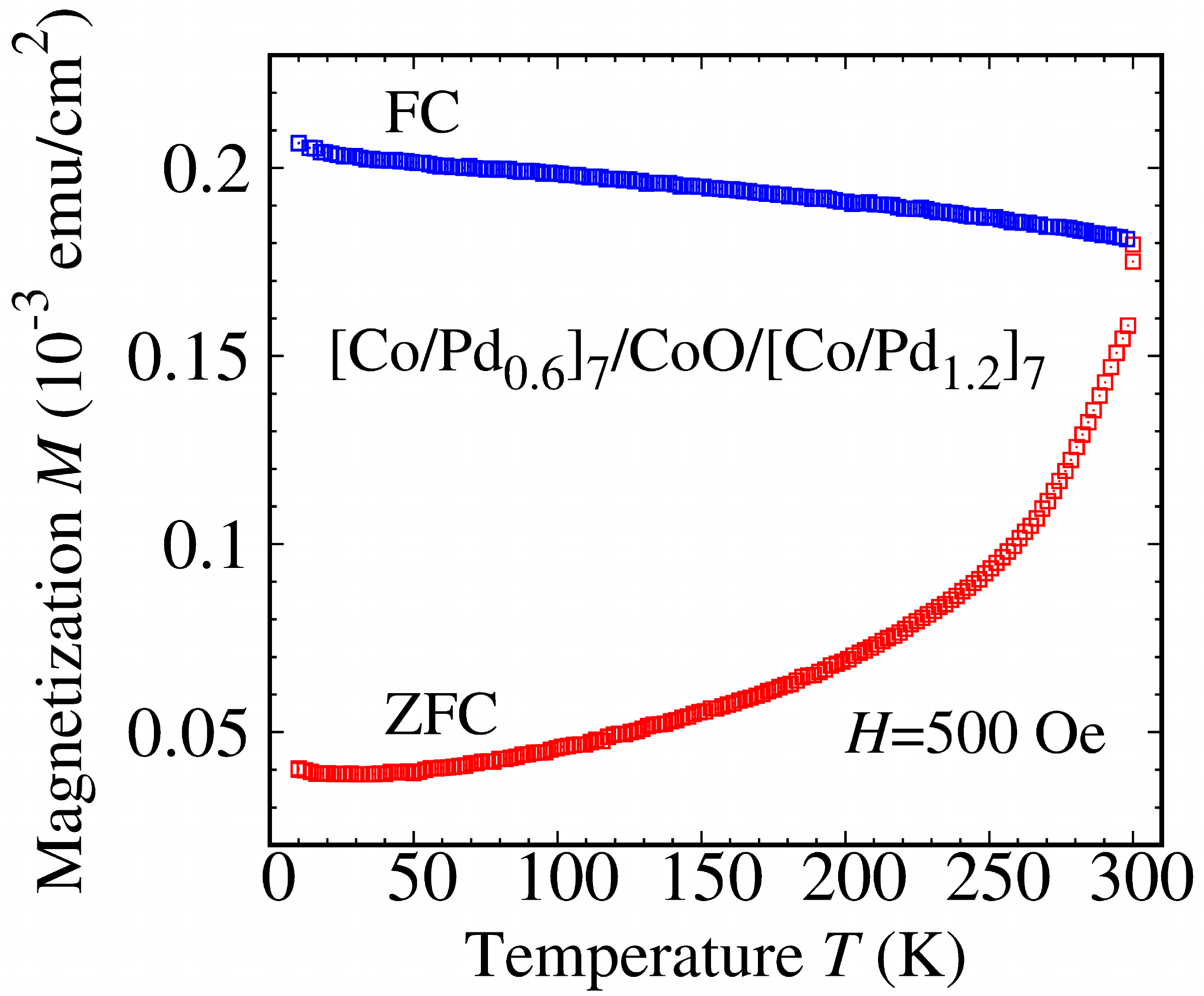}
 \caption{ZFC/FC curves for the [Co/Pd$_{\mathrm{0.6 \ nm}}$]$_{7}$/CoO/[Co/Pd$_{\mathrm{1.2 \ nm}}$]$_{7}$ composite measured with +500~Oe external magnetic field.}
 \label{Fig3}
\end{figure}

The zero-field-cooled (ZFC) and field-cooled (FC) curves for the SM/AFM/HM system are shown in Figure~\ref{Fig3}. 
The system was demagnetized at 300~K and, due to the cooling in zero magnetic field, that state was preserved at 10~K. 
Application of the 500 Oe external field induced nonzero magnetization caused by the partial orientation of the magnetic moments along the field direction. 
Heating of the system led to the increase of thermal fluctuations of the magnetic moments leading to more moments being unblocked and able to align with the field. 
This process can be observed in the ZFC curve as a~progressive rise of the magnetization up to 300~K. 
Cooling the system from 300~K (the FC curve) resulted in a~gradual decrease of thermal fluctuation energy, resulting in the blocking of the magnetic moments. 
On the other hand, the presence of the external magnetic field caused stepwise reorientation of the spins along the field direction. 
These two factors are responsible for the progressive increase of the magnetization seen in the FC measurement. 
The shape of the ZFC and FC curves is typical for ferromagnets and indicates that the composite reveals ferromagnetic behavior in the whole temperature range from 10~K to 300~K.
Since both curves are monotonic without any maxima, no superparamagnetic or superferromagnetic effects have to be taken into account during further analysis of the magnetization switching process.\cite{Per16AMI}
Additionally, in both FC and ZFC curves there is a~small increase of the signal at low temperatures. 
Such behavior suggests that a~small paramagnetic contribution is present in the system.

\begin{figure*}[th]
 \centering
 \includegraphics[width=0.75\textwidth]{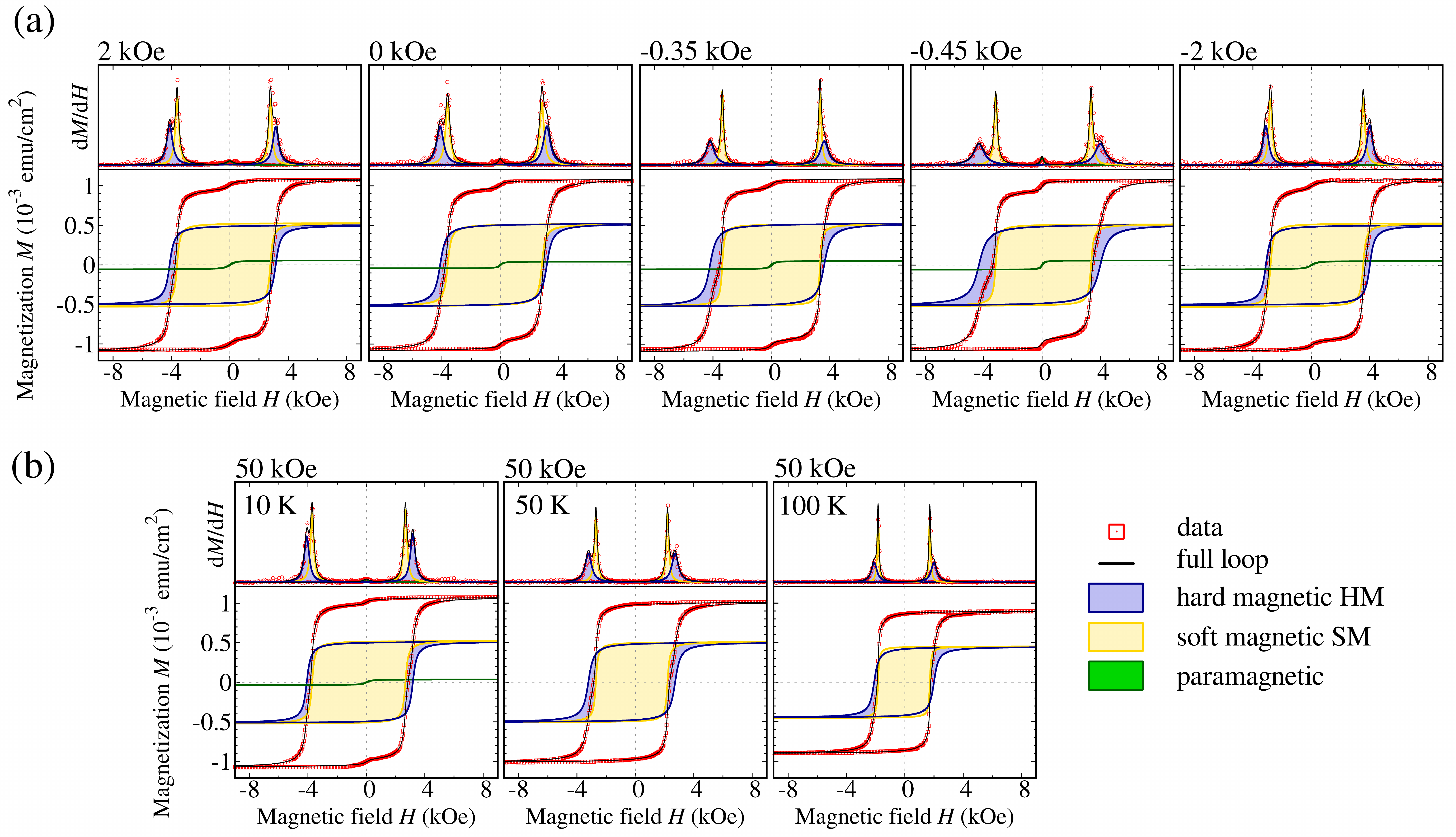}
 \caption{(a) Representative hysteresis loops measured for the [Co/Pd$_{\mathrm{0.6 \ nm}}$]$_{7}$/CoO/[Co/Pd$_{\mathrm{1.2 \ nm}}$]$_{7}$ composite at 10~K for different cooling fields. 
 (b) Hysteresis loops measured at 10 K, 50 K, and 100 K after field cooling in +50~kOe. 
 The upper panels in both figures show the d$M$/d$H$ switching field distributions.
 The points represent experimental data, and the solid lines are fits (see text). The hard (HM) and soft (SM) magnetization components are marked with corresponding color fields.}
 \label{Fig4}
\end{figure*}

The magnetization reversal mechanism of the exchange-biased SM/AFM/HM composite was studied by a~series of hysteresis loops measured at 10~K in the out-of-plane geometry. 
Prior to the measurement the external perpendicular magnetic field of +50~kOe was set at 300~K to align all ferromagnetic moments in the SM and HM layer in the positive out-of-plane direction. 
Then, the field was changed to $H_{\mathrm{cool}}$ in which the sample was cooled down to 10~K and the loops were measured. 
Representative hysteresis loops for various $H_{\mathrm{cool}}$ values are shown in Figure~\ref{Fig4}a. 
In all loops two pairs of switching fields per magnetization branch are present, similar to those observed for SM/HM system, indicating a~comparable spring-like behaviour of the composite film with SM and HM multilayers reversing at different external fields, lower and higher, respectively. 
Moreover, in the $M(H)$ loops are observed small signal steps around $H=0$, reflected in the d$M$/d$H$ curves as small maxima centered around zero external field.

Similarly to the previous cases, the upper and lower magnetization branches of the loops were fitted by the function of Eq.~\eqref{single_M_H}. 
However, here Eq.~\eqref{double_M_H} used for the SM/HM system was complemented by another component labeled as C to reflect the presence of the signal observed for $H\! \approx \!0$~Oe, giving the following expression
\begin{equation}
 \begin{split}
  M(H) = \frac{2}{\pi} M_{\mathrm{s}}^{\mathrm{SM}} \arctan \left( g^{\mathrm{SM}} \left[ \frac{H-H_{\mathrm{c}}^{\mathrm{SM}}}{H_{\mathrm{c}}^{\mathrm{SM}}} \right] \right) \\
  + \frac{2}{\pi} M_{\mathrm{s}}^{\mathrm{HM}} \arctan \left( g^{\mathrm{HM}} \left[ \frac{H-H_{\mathrm{c}}^{\mathrm{HM}}}{H_{\mathrm{c}}^{\mathrm{HM}}} \right] \right) \\
  + \frac{2}{\pi} M_{\mathrm{s}}^{\mathrm{C}} \arctan \left( g^{\mathrm{C}} \left[ \frac{H-H_{\mathrm{c}}^{\mathrm{C}}}{H_{\mathrm{c}}^{\mathrm{C}}} \right] \right) \ . 
 \end{split}
\label{triple_M_H} 
\end{equation}
The fits showed that the C component reveals no coercivity and its saturation magnetization $M_{\mathrm{s}}^{\mathrm{C}}$ accounts in average for 5\% of the total magnetization exhibited by the SM/AFM/HM system. 
Figure~\ref{Fig4}b demonstrates the evolution of the hysteresis loop for increased temperature indicating a~gradual decrease of the saturation magnetization as well as the reduction of the coercive field, accompanied by the reduction of the exchange bias field. 
Moreover, after heating the SM/SFM/HM systems to 50~K or 100~K the signal for $H=0$ does not appear. 
Taking into account the shape of the ZFC/FC curves (Figure~\ref{Fig3}) and the results obtained from the hysteresis loops it can be concluded that a~small fraction of a~paramagnetic phase is present in the SM/AFM/HM system.
This is due to the $M \! \propto \! T^{-1}$ Curie law according to which a~paramagnetic contribution to the magnetic signal becomes more prominent as temperature approaches zero.  
The origin of such phase can be related to the oxidation procedure applied to obtain antiferromagnetic CoO. 
Most of the Co volume deposited between the SM and HM stacks transformed into antiferromagnet. 
Therefore, the antiparallel arrangement of the magnetic moments gives zero net magnetization and does not contribute to the magnetization recorded in the measurements. 
However, there are still some Co atoms which were not oxidized and provide paramagnetic contribution to the system. 
Due to their magnetic nature they are not magnetically coupled to the other magnetic regions. 
Thereby, this phase does not have influence on the magnetic properties revealed by the soft, hard, and antiferromagnetic components of the system, and does not affect the reversal mechanism present in the composite.  

For each value of the cooling field $H_{\mathrm{cool}}$ the saturation magnetizations $M_{\mathrm{s}}^{\mathrm{SM}}$ and $M_{\mathrm{s}}^{\mathrm{HM}}$ of the soft and hard ferromagnetic components are approximately equal and similar to those observed for the SM and HM stacks (Figure~\ref{Fig1}a and b), and to the magnetization components obtained from fitting of the SM/HM hysteresis loop (Figure~\ref{Fig1}c).
Thereby, the overall magnetization of the SM/AFM/HM system, neglecting the paramagnetic contribution described above, is similar to that recorded for the SM/HM system, testifying that the magnetic signal comes only from the Co atoms within the [Co/Pd] stacks. 
The oxidized Co layer becomes antiferromagnetic below the N\'eel temperature, and preserves this property regardless of the cooling procedure, which is observed as a~lack of the overall magnetization change upon the $H_{\mathrm{cool}}$ alteration. 

\begin{figure*}[!h]
 \centering
 \includegraphics[width=0.50\textwidth]{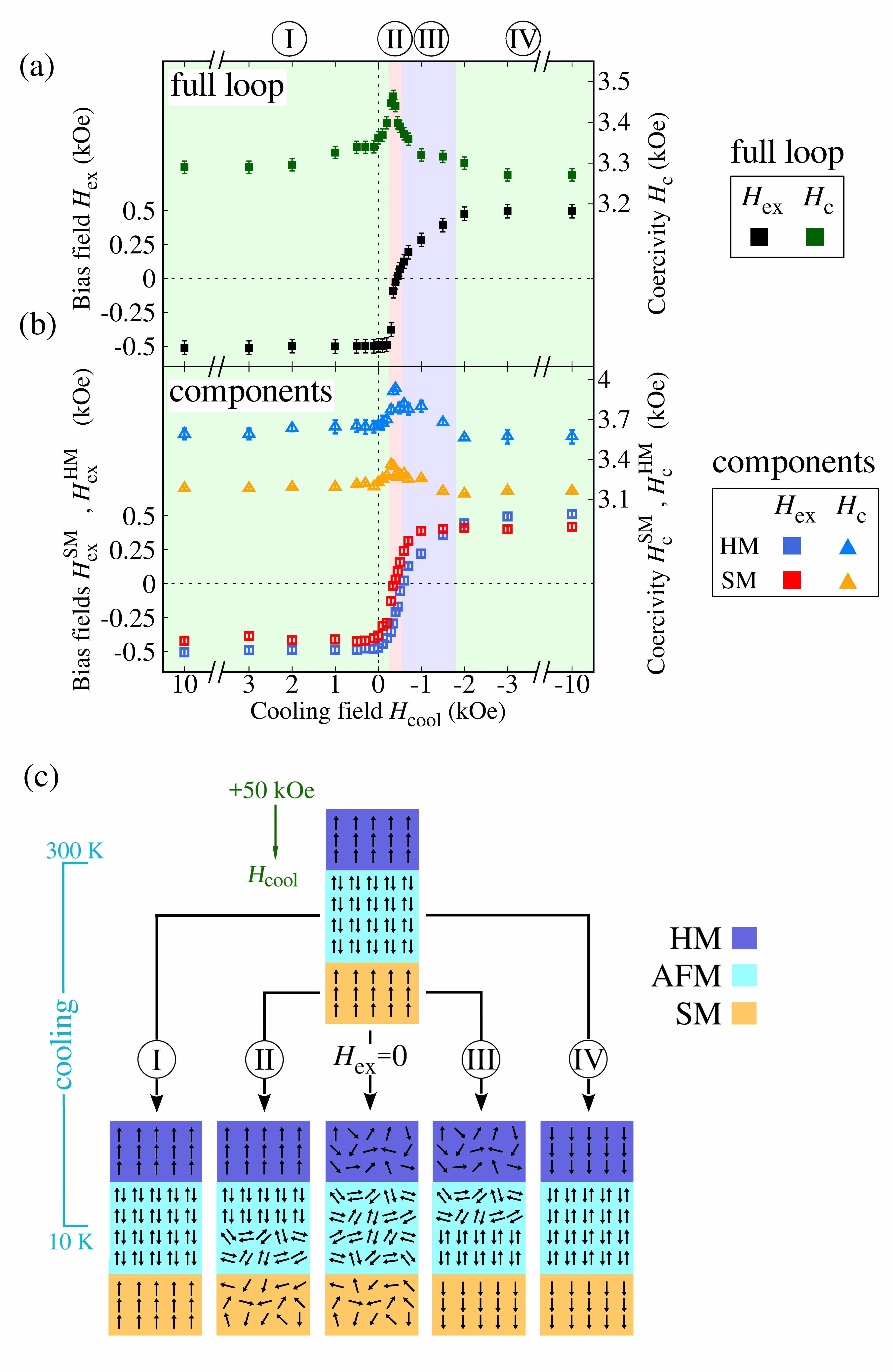}
 \caption{(a) Dependencies of the full loop exchange bias field $H_{\mathrm{ex}}$ (black squares) and coercivity $H_{\mathrm{c}}$ (green squares) on the cooling field $H_{\mathrm{cool}}$ for the SM/AFM/HM composite. (b) Dependencies of the exchange bias fields $H_{\mathrm{ex}}^{\mathrm{SM}}$ and $H_{\mathrm{ex}}^{\mathrm{HM}}$ (full squares) and coercivities $H_{\mathrm{c}}^{\mathrm{SM}}$ and $H_{\mathrm{c}}^{\mathrm{HM}}$ (full triangles) on the cooling field $H_{\mathrm{cool}}$ for the soft (SM) and hard (HM) magnetization components of the SM/AFM/HM composite. (c) Schematic representation of different magnetization reversal mechanisms for the different regions marked in subfigures (a) and (b) by Roman numbers.}
 \label{Fig5}
\end{figure*}

Changes of the exchange bias fields and coercivities on the cooling field $H_{\mathrm{cool}}$, obtained from the fits, are presented in Figures~\ref{Fig5}a and b. 
The dependencies can be divided into four regions in which the magnetization reversal process takes place in a~different manner. 

In the first region (I), covering the $H_{\mathrm{cool}}$ range from high positive fields down to -0.2~kOe, there is no significant change of the full loop bias field $H_{\mathrm{ex}}$, equal to -0.5~kOe. 
The corresponding $H_{\mathrm{ex}}^{\mathrm{SM}}$ and $H_{\mathrm{ex}}^{\mathrm{HM}}$ values also do not alter with the $H_{\mathrm{cool}}$, the $H_{\mathrm{ex}}^{\mathrm{SM}}$ is -0.4~kOe, and the $H_{\mathrm{ex}}^{\mathrm{HM}}$ is slightly larger than the $H_{\mathrm{ex}}$. 
The initial field of +50~kOe forces all magnetic moments in both SM and HM multilayers to align and to point in a positive out-of-plane direction. 
The absence of the $H_{\mathrm{ex}}$ changes suggests that in this region the $H_{\mathrm{cool}}$ field in this range is too weak to alter the original arrangement of the spins. 
Therefore, the system is cooled down with all ferromagnetic moments pointing in the direction perpendicular to the sample plane (see corresponding path I in Figure~\ref{Fig5}c). 
This lack of a~spin reorientation upon external field change, even for slightly negative $H_{\mathrm{cool}}$ values, is a~consequence of a~large out-of-plane anisotropy accompanied by the high magnetic remanence present in both SM and HM layers. 
Below the N\'eel temperature the CoO orders antiferromagnetically, and below the exchange bias blocking temperature its magnetic moments start to couple with the SM and HM ferromagnetic spins. 
Due to this exchange coupling the orientation of the ferromagnetic spins imposes the out-of-plane alignment of the AFM moments and freezes them, making the AFM spin structure insensitive to further changes of the external magnetic field. 
Such highly-ordered perpendicular arrangement of the SM, HM, and AFM magnetic moments favors a~strong out-of-plane exchange bias coupling on both SM/AFM and AFM/HM interfaces which is reflected as a~largest value of $H_{\mathrm{ex}}$, seen also for both SM and HM components. 

The second region (II) covers the $H_{\mathrm{cool}}$ range from -0.2~kOe to -0.5~kOe. 
Here, the $H_{\mathrm{ex}}$ value changes rapidly and alters its sign from negative to positive.  
As previously, the external field was set to +50 kOe and then changed to $H_{\mathrm{cool}}$, prior to the cooling and loop measurement. 
Contrary to the previous case, the cooling field becomes strong enough to drive the changes in ferromagnetic spin orientation and to start magnetization reversal process within the SM and HM layers. Since the SM has a lower coercivity we can expect that arrangement of its magnetic moments is more sensitive to the external field change. 
Additionally, we can anticipate that the spin arrangement in the SM layer diverges from the ideal out-of-plane orientation to a~greater extent than for the HM with larger coercivity (see path II in Figure~ \ref{Fig5}c).  
Therefore, at the AFM/HM interface the CoO spins are coupled antiferrmagnetically to each other and aligned mostly perpendicularly as imposed by the exchange coupling to the HM magnetic moments whose out-of-plane orientation is driven by the external field.
On the other hand, the random orientation of the SM magnetic moments induced by the cooling field caused formation of the variously oriented regions in the neighboring CoO layer.
Within these regions the AFM spins are coupled antiparallel and the spatial direction along which they are arranged is determined by the orientation of the nearby ferromagnetic moments due to the exchange interaction induced during the cooling procedure.   
Since the cooling field caused a~high degree of spin disorder in the SM layer the number of the perpendicularly oriented AFM magnetic moments is smaller than in the region (I). 
This results in the decrease of the out-of-plane exchange bias field $H_{\mathrm{ex}}$ and is also reflected in a~faster reduction of the $H_{\mathrm{ex}}^{\mathrm{SM}}$ value upon $H_{\mathrm{cool}}$ change than for the $H_{\mathrm{ex}}^{\mathrm{HM}}$ field (see Figure~\ref{Fig5}b).

Further change of the $H_{\mathrm{cool}}$ to the value of -0.45~kOe (see path in Figure~\ref{Fig5}c) leads to the bias field $H_{\mathrm{ex}}=0$. 
In this situation only a~small fraction of the HM spins is aligned in the initial positive out-of-plane direction. 
Thereby, this restricts the amount of the CoO material being frozen in the same orientation and causes a~meaningful reduction of the $H_{\mathrm{ex}}^{\mathrm{HM}}$ field measured in the out-of-plane direction. 
In the case of the SM layer, since it has lower coercivity than the HM, the cooling field was sufficiently large to rotate a~limited number of the magnetic moments and to point them in the negative out-of-plane direction. 
As a~consequence, the amount of the AFM spins coupled in the same manner is restricted, which leads to the appearance of poor positive out-of-plane $H_{\mathrm{ex}}^{\mathrm{SM}}$ bias field of the opposite sign 
in comparison to the $H_{\mathrm{ex}}^{\mathrm{HM}}$. 
The contributions from these two effects, taking place in the opposite spatial directions, compensate each other and give the overall exchange bias field $H_{\mathrm{ex}}$ equal to zero.

In the third region of the $H_{\mathrm{ex}}$ dependence, for the $H_{\mathrm{cool}}$ from -0.5~kOe to -2~kOe, the positive exchange bias field rises constantly but in a~less rapid way than in the second region. 
In this regime, it can be expected that the $H_{\mathrm{cool}}$ field is sufficiently large to reverse most of the spins in the SM layer and point them in the direction opposite to the initial state set by +50~kOe. 
However, the field is still not strong enough to fully rotate magnetic moments in the HM. 
Therefore, a~good out-of-plane arrangement of AFM spins extorted by the ordered ferromagnet is present mainly at the SM/AFM interface where the spins point in a~negative perpendicular direction. 
This leads to the perpendicular exchange coupling and gives rise to the positive bias loop shift (see accompanying path in Figure~\ref{Fig5}c).  
A~gradual change of the $H_{\mathrm{cool}}$ extorts an~out-of-plane spin alignment also in the HM. 
This results in an increase of the exchange coupling strength at the AFM/HM interface which, together with the coupling at the SM/AFM interface, is responsible for a~constant rise of the exchange bias field. 
The situation is similar to that observed in the second region with a~considerable difference in the $H_{\mathrm{ex}}$ slope. 
The magnetization reversal process in the SM is more abrupt than in the HM as it is seen in Figure~\ref{Fig1} for the [Co/Pd] multilayers. 
Therefore, when $H_{\mathrm{cool}}$ drives the magnetization reversal, a~small modification in the external field value can cause a~larger spin rotation in the SM than in the HM. 
Since the arrangement of the ferromagnetic moments is a~key parameter for the magnitude of the exchange bias field, the sharp reversal of the SM spins leads to a substantial change of $H_{\mathrm{ex}}$. 
The reversal of the HM requires larger $H_{\mathrm{cool}}$ fields and is more protracted, and for these reasons it provides less dynamic change of $H_{\mathrm{ex}}$ in the third region. 

\newpage
The fourth region extends from -2~kOe up to -3~kOe $H_{\mathrm{cool}}$ fields. 
The $H_{\mathrm{ex}}$ values are maximal and the same as in region I, but with the opposite sign, which indicates the same magnetization reversal mechanism. 
In the first region the $H_{\mathrm{cool}}$ field was too weak to change the orientation of the spins ordered by the initial +50~kOe field, so the spins were pointing in the positive out-of-plane direction. 
Here, the $H_{\mathrm{cool}}$ field is large enough to magnetically saturate the SM and HM layers in the opposite, negative, out-of-plane direction (see corresponding path in Figure~\ref{Fig5}c). 
Such orientation of the SM and HM moments maximizes their exchange coupling strength to the AFM producing the largest hysteresis loop shift. 

We observed that in region (I) the coercive field of the full loop has a~value of approximately 3.3~kOe. 
In region (II) it starts to increase, reaches its maximum for $H_{\mathrm{cool}}\!=\!-0.45$~kOe, and then decreases in region (III), reaching in region (IV) the same value as observed in region (I). 
Similar behavior is observed for the $H_{\mathrm{c}}^{\mathrm{HM}}$ and $H_{\mathrm{c}}^{\mathrm{SM}}$ coercive fields.     
In region (I), the magnetization reversal taking place along the upper hysteresis branch is more energetically demanding than the reversal along the lower branch due to the unidirectional anisotropy induced by the ferromagnetic-antiferromagnetic coupling.  
In region (IV) this situation is mirrored since the cooling field strength was sufficient to magnetically saturate both SM and HM components. 
Considering region (II), a~stepwise change of the cooling field leads to progressive spin orientation disorder induced in the ferromagnetic layers. 
The SM layer is more susceptible to this disarranging process due to its softer ferromagnetic properties, contrary to the HM component revealing larger coercitivy. 
As a~result, the exchange coupling process causes formation of the variously spatially oriented regions within the antiferromagnetic layer below the blocking temperature during the cooling procedure. 
These regions, maintaining their antiferromagnetic nature, couple the ferromagnetic layers in diverse spatial directions and constrict the reversal process along both upper and lower magnetization branches.  
Since that, the reversal process becomes more energetically demanding as the spatial orientations of the pinning AFM regions get more diverse, leading to the coercivity rise.  
For the cooling field of -0.45~kOe, at which the coercivity is maximal, some of the regions in the antiferromagnet at the SM/AFM interface extort already the appearance of small positive exchange bias field, while the bias field associated with the AFM/HM interface still remains negative. 
This means that some of the AFM regions counteract against the reversal along the upper magnetization branch, due to the pinning effects driven by the exchange coupling, while the others constrict the magnetization switching in the opposite field sweep direction. 
Coexistence of these two processes causes the reversal is energetically demanding along both branches, leading to the increase of coercivity. 
In region (III) both $H_{\mathrm{ex}}^{\mathrm{HM}}$ and $H_{\mathrm{ex}}^{\mathrm{SM}}$ bias fields changed their sign to positive. 
In this situation, a~stepwise change of the external cooling field extorts gradual spin ordering along the negative perpendicular direction.
Thereby, the magnetization reversal along the upper hysteresis branch becomes less energetically demanding than in region (II) due to the fact that the exchange anisotropy at the SM/AFM and AFM/HM interfaces tends to couple the ferromagnetic layers along the same direction. 
At the same time, since the cooling field orientation induced the unidirectional anisotropy in the opposite direction than in region (I), the reversal along the lower branch becomes less difficult.  
This results in the progressive decrease of the coercivity to the minimal value observed in regions (I) and (IV).

To support the information on the magnetization reversal mechanism provided by the analysis of the $H_{\mathrm{ex}}$ and $H_{\mathrm{c}}$ dependencies on the cooling field $H_{\mathrm{cool}}$, a~series of FORC measurements were carried out. 
The experiments were done at 10~K for the SM/AFM/HM system cooled in the external fields of $+50$~kOe and $-0.45$~kOe to obtain FORC distributions for the cases of maximal exchange bias field value, and for no biased loop. 
Materials with the exchange bias effect show training effect which is observed as a~gradual decrease of the exchange bias field magnitude through consecutive hysteresis loop cycling.\cite{Bin04PRB} 
The $|H_{\mathrm{ex}}|$ falloff is mainly present during the first few cycles, and after that the changes of the exchange field become less prominent. 
During the FORC measurement the external field is switched many times between the reversal $H_{\mathrm{R}}$ and saturation $\pm10$~kOe fields.
Therefore, starting the measurement from the untrained state would be associated with the presence of the training effect affecting the resultant distribution, especially for a~few initial $M(H_{\mathrm{R}},H)$ curves.
To minimize the influence of this effect on the final FORC, prior to the measurements at low temperature, the external magnetic field was switched from +50~kOe to -50~kOe 15 times.
\begin{figure*}[th]
 \centering
 \includegraphics[width=0.65\textwidth]{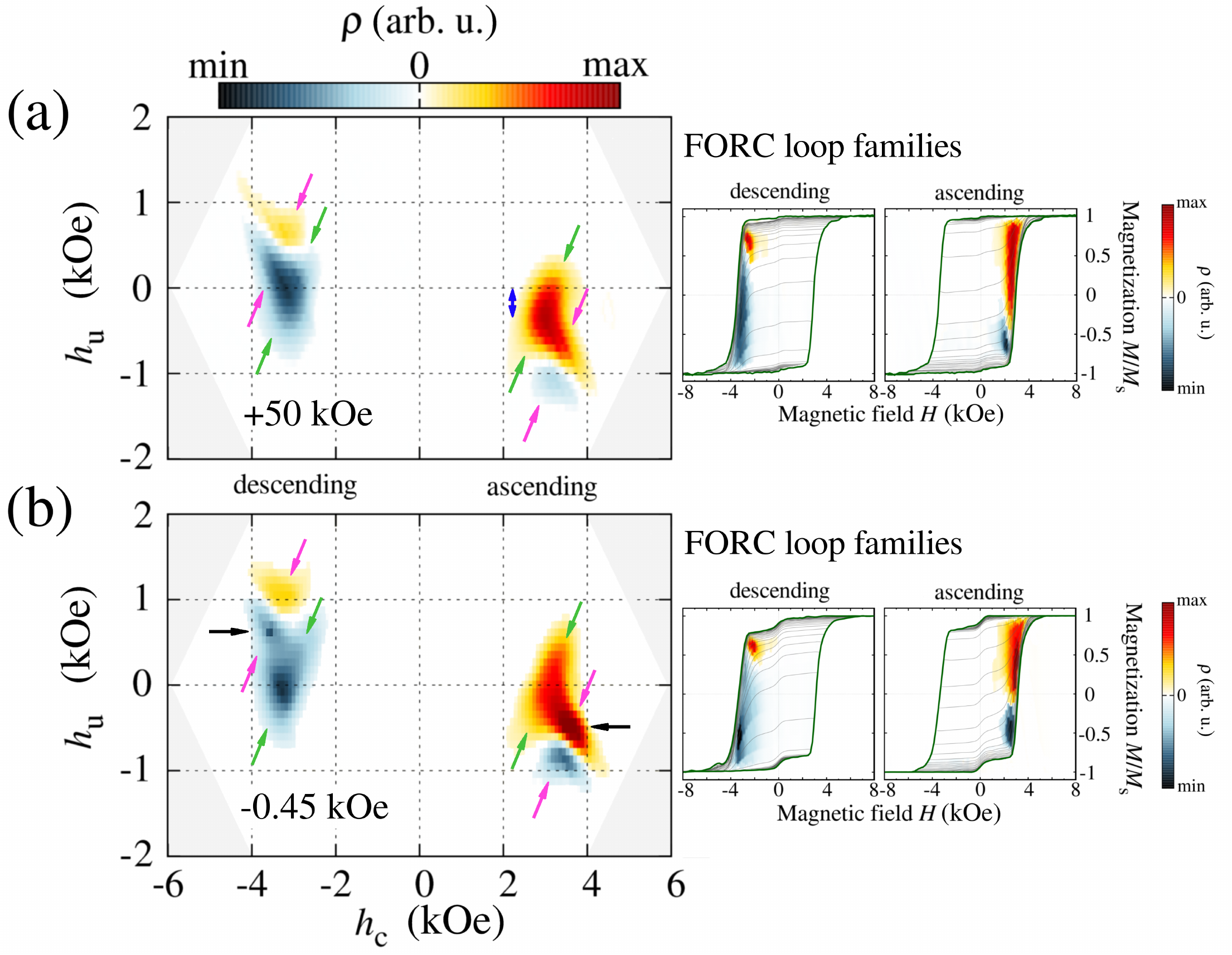}
 \caption{FORC measurements for the [Co/Pd$_{\mathrm{0.6 \ nm}}$]$_{7}$/CoO/[Co/Pd$_{\mathrm{1.2 \ nm}}$]$_{7}$ composite obtained at 10~K for different $H_{\mathrm{cool}}$ cooling fields: (a) +50~kOe, and (b) -0.45~kOe. Left: $M(H,H_{\mathrm{R}})$ FORC loop families for ascending and descending magnetization branches, plotted in canvas of the hysteresis loops. The characteristic features are marked with color arrows (see text).}
 \label{Fig6}
\end{figure*}

The initial magnetization data dependent on the reversal field $H_{\mathrm{R}}$ and the external field $H$, measured for gradual ascending and descending $H$ and $H_{\mathrm{R}}$, were transformed into FORC distributions $\rho(H_{\mathrm{R}},H)$ by calcutextitg mixed second-order derivatives of the magnetization $M(H_{\mathrm{R}},H)$:\cite{Pal16N}
\begin{equation}
 \rho(H_{\mathrm{R}},H) = - \frac{1}{2} \frac{\partial^{2} M(H_{\mathrm{R}},H)}{\partial H \partial H_{\mathrm{R}}} \ . 
\end{equation}
Such a~definition of the FORC $\rho(H_{\mathrm{R}},H)$ distribution eliminates reversible components of the magnetization reversal process. 
Therefore, any nonzero $\rho(H_{\mathrm{R}},H)$ signal represents an irreversible magnetization switching. 
The FORC $\rho$ distribution can be plotted in coordinate system rotated by 45$^{\circ}$ with the coercive field $h_{\mathrm{c}}$ and interaction field $h_{\mathrm{u}}$ on the horizontal and vertical axes, respectively. 
These coordinates are defined as follows:
\begin{equation}
 h_{\mathrm{c}} = \frac{H - H_{\mathrm{R}}}{2} \ , \  h_{\mathrm{u}} = \frac{H + H_{\mathrm{R}}}{2} \ .
\end{equation}
In a~such coordination system, $h_{\mathrm{c}}$ represents the local coercive field of each hysteron, and $h_{\mathrm{u}}$ corresponds to the local interaction field shifting the hysteron's center along the field axis.\cite{Rut17SR} 
The obtained FORC distributions $\rho(h_{\mathrm{c}},h_{\mathrm{u}})$ for both ascending and descending magnetization branches, together with FORC curve families $M(H_{\mathrm{R}},H)$ plotted in canvas of the hysteresis loop, are shown in Figure~\ref{Fig6}. 
In each FORC distribution there are two main prominent features related to the irreversible magnetization switching processes. 
These features are characteristic of the reversal process taking place by the nucleation of the magnetic domains and subsequent domain-wall motion.\cite{Rah09APL}

First, the ridges seen in all FORCs and marked by the green arrows in Figure~\ref{Fig6} are directly related to the nucleation and subsequent rapid propagation of the magnetic domains with reversed magnetization orientation within the system. 
It can be seen in both FORCs measured for descending magnetization branches that the ridge is centered on the point with no shift along the vertical $h_{\mathrm{u}}$ axis. 
On the other hand, the ridge centre observed for ascending fields after cooling the sample in +50~kOe is moved along $h_{\mathrm{u}}$ for approximately -0.3~kOe (blue arrow in Figure~\ref{Fig6}a). 
This feature is connected to the presence of the exchange interaction in the system which is, in our case, the ferromagnetic-antiferromagnetic coupling present at the SM/AFM and AFM/HM interfaces in the composite. 
The value of the $h_{\mathrm{u}}$ shift is lower than the exchange bias field -0.5~kOe observed in the hysteresis loop. 
However, it has to be taken into account that the FORCs were measured on trained system where the value of the $H_{\mathrm{ex}}$ is expected to be lower than for untrained composite. 
Therefore, the $h_{\mathrm{u}}$ shift is the evidence for the presence of ferromagnetic-antiferromagnetic coupling in the system which drives the exchange bias effect. 
For the case of $H_{\mathrm{cool}}\!=\!-0.45$~kOe when no hysteresis loop shift was recorded (see Figure~\ref{Fig5}) the corresponding ascending ridge also reveals no shift along the interaction axis~$h_{\mathrm{u}}$.

The second feature observed in each FORC is a~positive-negative pair of irreversible regions along the lines marked by the purple arrows in Figures~\ref{Fig6}a and~b. 
Such paired signals are distinctive for switching of the orientation of the magnetization by domain nucleation and domain-wall motion, and are related to the annihilation of magnetic domains at the end of the magnetization reversal process as the external field induces magnetic saturation of the system.
It is also seen in the FORCs for the system cooled in $H_{\mathrm{cool}}\!=\!-0.45$~kOe (Figure~\ref{Fig6}b) that the irreversible regions related to the domain annihilation are better pronounced than the corresponding features observed after field cooling in +50~kOe (black arrows in Figure~\ref{Fig6}b).

All FORCs show the lack of irreversible signal around $h_{\mathrm{c}}\!=\!0$. 
This confirms that the magnetization drop observed in the hysteresis loops for external magnetic field around 0 (see Figure~\ref{Fig4}a) has a~paramagnetic nature. 
The paramagnetic contribution is also reflected in the FC and ZFC curves (Figure~\ref{Fig3}) and observed as a small magnetization rise at a~low temperature. 
The origin of the effect is the presence of unoxidized Co atoms in the AFM. 
They do not exhibit ferromagnetic behavior and are not magnetically coupled to the other magnetic regions. 
Due to their nature they do not influence the magnetization reversal process of the composite.      

\newpage
\vspace{-0.25cm}
\section{Conclusions}

\vspace{-0.15cm}
In this paper, we describe the magnetization reversal mechanism of the exchange-biased [Co/Pd$_{\mathrm{0.6 \ nm}}$]$_{7}$/CoO/[Co/Pd$_{\mathrm{1.2 \ nm}}$]$_{7}$ composite where the antiferromagnetic film is sandwiched between the soft and hard ferromagnetic layers. 
Upon cooling such an exchange-spring-like system in the external magnetic field $H_{\mathrm{cool}}$ the initial alignment of the ferromagnetic spins forces the antiferromagnet to freeze in a~certain state. 
This state, induced by the cooling field, is then crucial for the magnetization switching process, which has been reflected as the change of the magnitude and the sign of the exchange bias field $H_{\mathrm{ex}}$ with the $H_{\mathrm{cool}}$. 
Since we have two different ferromagnetic films from both sides of the AFM, the shape of the exchange bias field $H_{\mathrm{ex}}$ of the cooling field $H_{\mathrm{cool}}$ dependence is conditioned by the reversal mechanisms manifested by these two materials. 
In case where the external field is large enough to affect the spin orientation in the soft magnet, simultaneously being too weak to rotate moments of the hard magnet, we observed rapid change of the exchange bias field. 
On the other hand, the rotation of the magnetic moments of the hard magnetic film induced by the $H_{\mathrm{cool}}$ resulted in less prominent slope of the $H_{\mathrm{ex}}$ curve. 
Therefore, the dependence is not symmetrical around the point $H_{\mathrm{ex}}\!=\!0$. 
We also observed that cooling the system in zero field does not result in the lack of the exchange bias loop shift, which is caused by large out-of-plane anisotropy, leading to high magnetic remanence displayed by the system. 
The FORC studies show that the magnetization reversal is driven by the nucleation of magnetic domains with opposite orientation followed by the domain-wall motion. 
Additionally, we record the shift of the FORC irreversible region caused by the presence of the exchange bias field in the system.

\bibliographystyle{plainnat}
\vspace{1.1ex}
\begin{center}
 $\star$ $\star$ $\star$
\end{center}

\vspace{-9ex}

\setlength{\bibsep}{0pt}
\renewcommand{\bibnumfmt}[1]{$^{#1}$}

\end{document}